\documentclass[square,aip,preprint,showkeys,superscriptaddress]{revtex4}
\usepackage{graphicx}
\usepackage{epstopdf,amsmath}

\begin{document}

\title{Electrical conduction mechanisms of metal / high-T$_c$ superconductor (YBCO) interfaces}

\author{L. F. Lanosa}
\thanks{Present address: Departamento de Matem\'atica - FCEyN - Universidad de Buenos Aires and IMAS, UBA-CONICET, Pabell\'on I, Ciudad Universitaria, C1428EHA Buenos Aires, Argentina}
\affiliation{Laboratorio de Bajas Temperaturas, Departamento de
F\'{\i}sica, FCEyN, Universidad de Buenos Aires and IFIBA,
UBA-CONICET, Pabell\'on I, Ciudad Universitaria, C1428EHA Buenos
Aires, Argentina}
\author{H. Huhtinen}
\affiliation{Wihuri Physical Laboratory, Department of Physics,
University of Turku, FI-20014 Turku, Finland}
\author{P. Paturi}
\affiliation{Wihuri Physical Laboratory, Department of Physics,
University of Turku, FI-20014 Turku, Finland}
\author{C. Acha}
\thanks{corresponding author (acha@df.uba.ar)}
\affiliation{Laboratorio de Bajas Temperaturas, Departamento de
    F\'{\i}sica, FCEyN, Universidad de Buenos Aires and IFIBA,
    UBA-CONICET, Pabell\'on I, Ciudad Universitaria, C1428EHA Buenos
    Aires, Argentina}

\date{\today}


\begin{abstract}
Current-voltage characteristics of Au~/~YBa$_2$Cu$_3$O$_{7-\delta}$
interfaces (Au/YBCO), built on optimally-doped YBCO thin films,
grown by pulsed laser deposition, were measured as a function of
temperature in the 50 K to 270 K range, for two different resistance
states. A non-trivial equivalent circuit model is proposed, which
reveals the existence of a highly inhomogeneous scenario composed by
two complex layers: one presenting both a non-linear Poole-Frenkel
conduction as well as Variable Range Hopping localization effects
(probably associated with YBa$_2$Cu$_3$O$_{6}$) mixed with a minor
metallic phase, while the other is also composed by a mixture of
YBCO with different oxygen contents, where a metallic ohmic phase
still percolates. A microscopic description of the effects produced
by the resistance switching is given, showing the evolution of
carrier traps, localization effects and dielectric behavior for each
state. The dielectric behavior is interpreted in terms of a
Maxwell-Wagner scenario.

\end{abstract}



\maketitle

\section{INTRODUCTION}

In the last years, memory devices based on the resistive switching
mechanism (RS), usually called RRAM, have focused great interest of
the scientific and technological community due to the potential
technological interest in terms of their application on denser
non-volatile memories, with lower energy consumption and better
resistance to hostile environments. Additionally, their capacity to
reproduce logical gates~\cite{Borghetti10} and to simulate synaptic
connections~\cite{Jo10} opened interesting paths towards the area of
electronic circuits, particularly to those that mimic neural
networks~\cite{Alibart13} and even reproduce the electrical behavior
of neurons.~\cite{Stoliar17}

Although great advances have been made to understand the physics
behind the memory properties of RRAM devices, as for example, the
explanation of the non-trivial behavior observed in the resistance
hysteresis loops of metal-transition metal oxide bipolar devices by
a voltage-enhanced oxygen diffusion model~\cite{Rozenberg10}, many
aspects are still not addressed in order to fully understand the
singular properties of these devices. In particular, considering the
oxygen migration easiness in these oxides, the existence of defects
or inhomogeneities in a real interface is highly probable and may
produce a significant effect on their transport properties. Many
factors may contribute to that non-uniform spatial oxygen
distribution near the interface, which may regulate the final oxygen
concentration distribution: the chemical oxygen affinity of the
metallic layer, deposited on top of the oxide surface with a certain
roughness, as well as the competition between oxygen
electromigration, favored in the high resistance zones, and the
oxygen diffusion, due to the Fick's law. Within this framework, the
knowledge of which are the microscopic factors that determine the
electrical conduction properties through these inhomogeneous
interfaces can then be exploited by material engineering strategies
in order to take full advantage of the resistance switching
properties of these devices.

Devices based on metal~/~YBCO (YBa$_2$Cu$_3$O$_{7-\delta}$) probably
will not be useful for massive technological applications due to
their not obvious integration to the Si-based electronics and to
their partial retentivity~\cite{Schulman12}, associated with their
high oxygen diffusivity in certain crystallographic
directions~\cite{Placenik12}. Instead, their interest is based in
their capacity to shed light on the electrical transport mechanisms
through a metal-complex oxide interface, which is a common feature
of many memristive interfaces and particularly to improve our knowledge of the electrical transport and dielectric properties of YBCO. Indeed, metal~/~YBCO devices have
shown interesting bipolar resistive switching
properties~\cite{Acha09a,Acha09b,Placenik10,Acha11,Schulman13}, with
particular relaxation effects~\cite{Schulman11,Placenik12}, the electrochemical control of YBCO's carrier density~\cite{Palau18} and
present some particular characteristics proper of inhomogeneous
interfaces.~\cite{Schulman12,Schulman15,Waskiewicz15,Truchly16,Waskiewicz18,Tulina18}.

In this paper, our objective is to identify the relevant transport
mechanism through the Au~/~YBCO interface by analyzing its
current-voltage (I-V) characteristics at different temperatures, in
order to determine its equivalent circuit model and to reveal the
temperature dependence of the microscopic factors that regulate each
resistive state.

Our results indicate the existence of a mixture of conduction
mechanisms, where the non-linearities, as previously
reported~\cite{Schulman12,Schulman15}, come from a Poole-Frenkel
emission, dominated by carrier traps on YBCO, in parallel with a
variable range hopping ohmic conduction, both in series with a
metallic conduction. These results reveal an scenario were the
interfacial zone is highly inhomogeneous in terms of the oxygen
distribution, leading to a mixture of conducting and isolating
zones. In this way, the elements that may produce a colossal
dielectric constant are present, as a
consequence of the Maxwell-Wagner effect.~\cite{VonHippel62,Lunkenheimer10}

\section{EXPERIMENTAL DETAILS}

Near optimally-doped and fully relaxed YBCO thin films were grown by pulsed
laser deposition (PLD) on top of a (100) single crystal STO
substrate. The films were deposited by applying 1500 pulses with a growth rate of 0.1 nm/pulse, producing the 150 nm thick layer as earlier confirmed by transmission electron microscopy (TEM) calibrations under the same deposition conditions.~\cite{Khan19}  Details of their synthesis and characterization can be found elsewhere (see also Suppl. Material) .~\cite{Huhtinen01,Paturi04,Peurla07}

\begin{figure}
    \vspace{0mm}
    \centerline{\includegraphics[angle=0,scale=0.7]{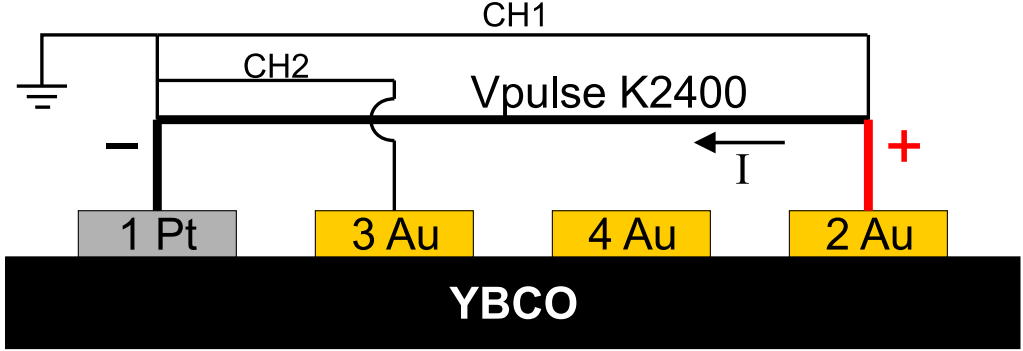}}
    \vspace{-5mm}\caption{(Color online) The YBCO thin film device with its electrodes and
        experimental configuration. Au was sputtered on pads $2$-$3$-$4$ while Pt on pad $1$.
        Pulses were applied between electrodes $"1"(-~polarity)$ and $"2"(+~polarity)$ with a
        Keithley 2400 SMU, while a Tektronix 3034 oscilloscope registered voltage drops between
        pads $2-1$ in Channel~1 and $3-1$ in Channel~2.} \vspace{-0mm}
    \label{fig:muestra}
\end{figure}

Metal-YBCO devices were prepared by using stencil lithography to
sputter four metallic electrodes on top of the YBCO thin films. The
sputtered electrodes have a width of 30 nm, a 0.7 x 0.7 mm$^2$ area
and a mean separation of 0.4 mm. Gold leads were carefully fixed
over them by using silver paint without contacting directly the
surface of the YBCO sample. Three of the electrodes were made with
Au, while the other with Pt, as depicted in Fig.~\ref{fig:muestra}.
As previously shown for YBCO ceramic slabs~\cite{Schulman13} and
checked for the YBCO thin films, this particular choice was based in
the fact that the Pt/YBCO interfaces have a lower resistance value
than the Au/YBCO ones ($R_{Pt}\lesssim R_{Au}/3$), as well as a
small RS amplitude. In this way, we may disregard the influence of
the Pt/YBCO electrode and proceed as if only the Au/YBCO electrode
is active (i.e. presents a relevant RS effect), simplifying the
effects produced upon voltage pulsing treatments. Thus, the studied
devices correspond to a Pt/\textbf{YBCO/Au} combination, arranged in
a planar structure. Pulses were applied between the Pt and Au
electrodes, labeled $"1"$ and $"2"$, respectively. The design
allowed 3-Wire (3W) and 4-Wire (4W) measurements using the remaining
electrodes $3$ and $4$: 3W to characterize individual metal/YBCO
interfaces, and 4W only for the YBCO characterization.

\begin{figure}
    \vspace{0mm}
    \centerline{\includegraphics[angle=0,scale=0.4]{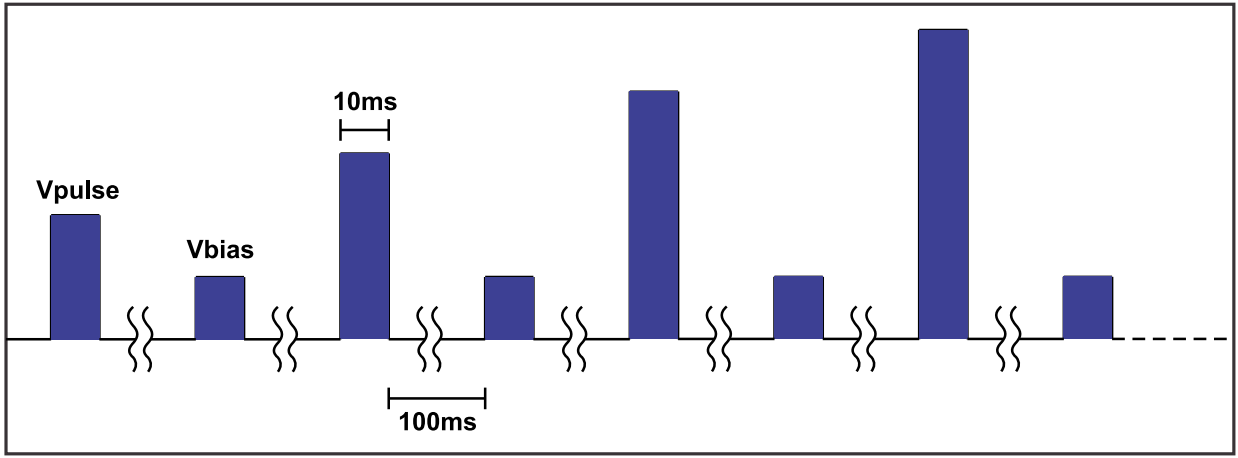}}
    \vspace{-5mm}\caption{Diagram of the voltage pulsing protocol used to study the effect of RS
        on the microscopic parameters that controls the electrical transport properties of the Au/YBCO
        interface. 10ms pulses ($V_{PULSE}$) were applied from 0 V to 5 V in 0.1 V steps, with
        an additional low voltage and fixed 0.1 V pulse ($V_{BIAS}$) to control and measure the remnant
        resistance ($R_{REM}$). Positive polarity pulses were applied for the characterization of
        the LRS, while negative ones for the HRS.} \vspace{-0mm}
    \label{fig:protocolo}
\end{figure}

In order to measure the I-V characteristics, 10 ms voltage pulses of
increasing amplitude were applied with a Keithley 2400 SMU between
electrodes $"1"$ and $"2"$, Pt($-$) and Au($+$), respectively. This
instrument also measure the circulating current (I) during the
pulses.  A $V_{BIAS}=0.1V$ pulse is also applied between pulses to
control and measure the remnant resistance, $R_{REM}$. The sketch of
the pulsing protocol can be observed in Fig.~\ref{fig:protocolo}.

At the same time a Tektronix 3034 oscilloscope was used to control
the integrity of the applied pulses (Channel~1), and to measure the
voltage drop between the Pt electrode $"1"$ and the intermediate Au
$"3"$ one (Channel~2). This 3W configuration served to determine the
voltage drop in the active $"2"$ Au/YBCO interface for studying its
I-V characteristics.

During all the I-V measurements, in order to characterize the same
resistive state at different temperatures, the voltage range was
carefully limited to avoid the RS of the device. To confirm this, we
checked that there were no sudden jumps in $R_{REM}$ during the
pulsing protocol.

As shown previously on ceramic YBCO/metal
interfaces~\cite{Schulman13}, the thin film YBCO/metal interface
presents a bipolar RS (see Suppl. Material). Briefly, after a
certain positive applied pulse [$V_{SET}\sim$~$5~V~@~270~K$,
polarity arbitrarily defined ($-$)Pt/YBCO/Au($+$)], the active
Au/YBCO interface sets to a low resistive state (LRS), while after a
negative one ($V_{RESET}\sim$~$-5~V~@~270~K$) sets to a high
resistive state (HRS).

For each I-V characteristic, temperature was stabilized within a 100
mK range for each selected temperature, varied from 270 K to 50 K at
$\sim$5 K intervals. Devices were located in a liquid Helium
cryostat where temperature was measured using a carbon-glass
thermometer, in good thermal contact with the device.

In this way, the device was initially set to the LRS at 270 K
($R^{LRS}_{REM} \simeq$ 4 k$\Omega$). Then its I-V characteristics
were measured twice at all the programmed temperatures (on cooling
and heating), obtaining similar results in both cases. After
stabilizing the temperature at 270 K, the sample was RESET to its
HRS ($R^{HRS}_{REM} \simeq$ 30 k$\Omega$) and the same procedure
followed to measure the I-V characteristics as a function of
temperature for the LRS was completed for the HRS.

The I-V characteristics obtained in the LRS are in the
$0V\rightarrow5V$ range, while in the HRS are in $-5V\rightarrow0V$.
As mentioned, the negative (positive) side in the LRS (HRS) could
not be measured in a relevant voltage range in order to avoid a RS.

\section{RESULTS AND DISCUSSION}

\begin{figure}[!htb]
    \centering
    \includegraphics[scale=.4]{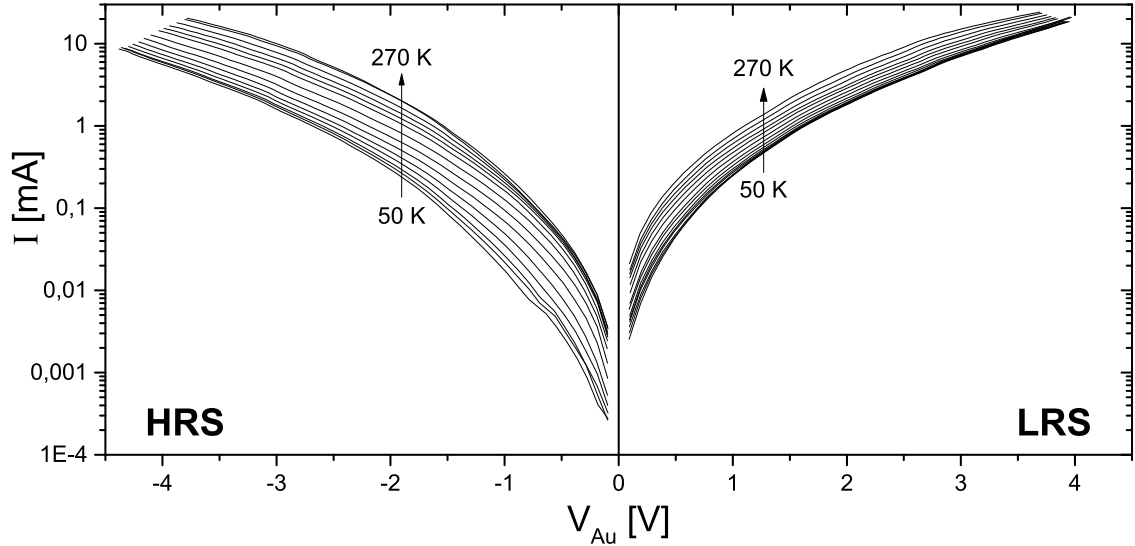}
    \caption{I-V characteristics of the Au-YBCO interface
 at different temperatures from 50 K to 270 K
        for a low (LRS) and a high (HRS) resistance state. For clarity,
        only curves measured increasing temperature with a $\simeq$ 15 K
        interval are shown.}
    \label{fig:IVs}
\end{figure}

The measured I-V characteristics as a function of temperature for
both states (LRS and HRS) can be observed in Fig.~\ref{fig:IVs}.

\noindent A systematic increase of the conductivity is observed when
increasing the temperature for both states. In the interest of
determining if these I-V curves can be associated with a typical
interfacial conduction mechanism~\cite{Sze06,Chiu14} [i.e. ohmic,
Schottky (Sch), Poole-Frenkel (PF), Space Charge Limited Currents
(SCLC), Fowler-Nordheim (FN), or to an eventual combination of
them], the graphical representation of the I-V curves based on the
power exponent $\gamma=dLn(I)/dLn(V)$ plotted as a function of
$V^{1/2}$, can be considered.~\cite{Acha17} In fact, this method has
proven to be extremely useful in determining the presence of
different transport mechanisms when more than one contributes to
electric transport.~\cite{Acha16,Acevedo17,Ghenzi19}

\begin{figure}[!htb]
    \centering
    \includegraphics[scale=.6]{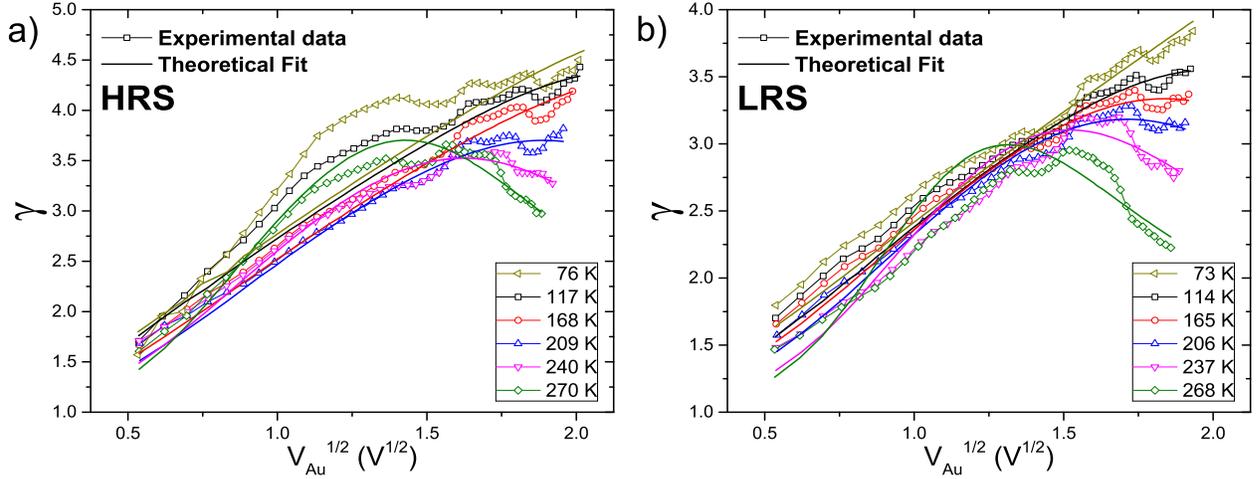}
    \caption{Experimental and theoretical (see Eq.~\ref{eq:I-V}) $\gamma$ representation of the I-V characteristics for the a) HRS and b) LRS.}
    \label{fig:gammaHRSyLRS}
\end{figure}

As shown in Fig.~\ref{fig:gammaHRSyLRS}, despite the noise
associated with the calculation of the derivative, it can be
observed that $\gamma$ follows a complex curve, indicating the
presence of more than one conduction process. For voltages tending
to cero, $\gamma$ extrapolates to values close to 1 (ohmic), while
for intermediate voltages, $\gamma$ follows a linear-like dependence
(Sch or PF), reaching a maximum, with a clear decrease for the
higher voltage range (ohmic or SCLC). The similarity of the curves
for both states indicates that the involved conduction mechanisms
remain present independently of the RS, which only changes the
relative weight of them.

\begin{figure}[!htb]
    \centering
    \includegraphics[scale=.5]{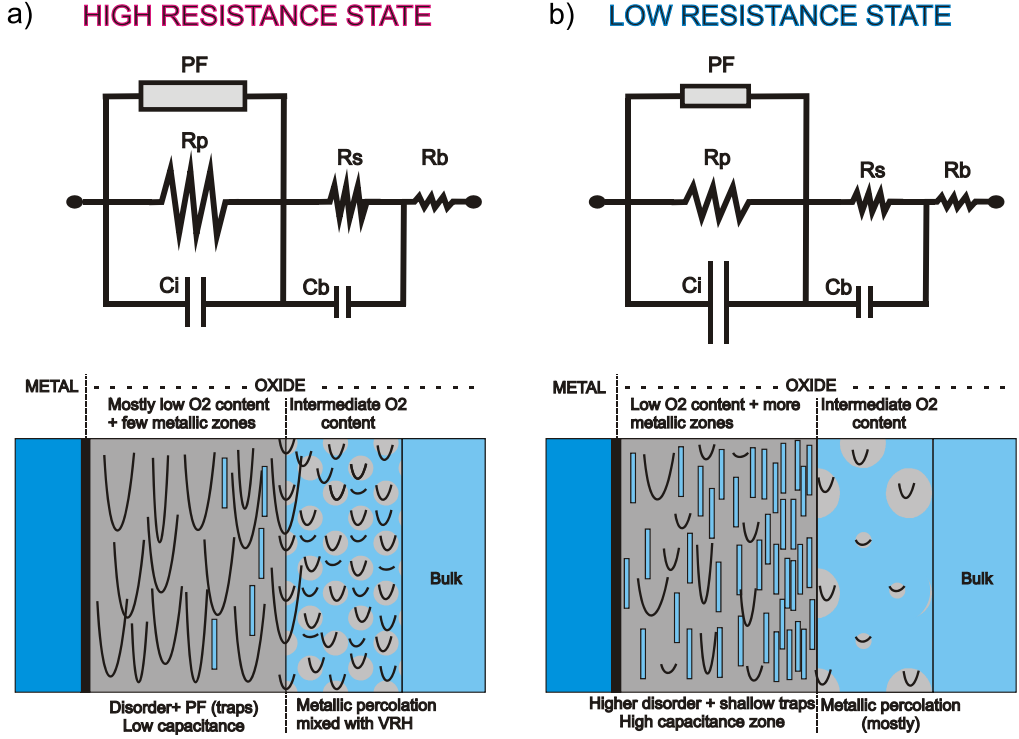}
    \caption{Schematic of the proposed equivalent circuit model for the Au/YBCO
    interface at a) HRS and b) LRS. The circuit includes a PF element in parallel with an ohmic
    resistor (R$_P$), both in series with a second ohmic resistor (R$_S$). In this
    way (see the discussion in the text), the interface is represented by 2
    oxygen-depleted regions: the first one, in contact with the metal, mostly composed by YBCO6,
    where oxygen vacancies act as traps and produce a disordered potential for the
    electrical carriers (PF+VRH), and the second one, with a lower vacancy content,
    where the metallic bulk phase still percolates. Notice that the first region also
    present conducting islands, whose density increases in the LRS, favoring the
    appearance of Maxwell-Wagner effects. }
        \label{fig:circuito}
\end{figure}

As no appreciable rectification was measured a Sch interface can be
ruled out. This is in accordance to the fact that ohmic contacts are
expected in interfaces where metals with high work function are
deposited on p-type oxides.~\cite{Batra91} In this way, the most
simple equivalent circuit that can be proposed is the one where the
non-linear behavior comes from a PF effect, associated with the
existence of traps within the oxide, in parallel with a leaky ohmic
conduction (R$_P$), both in series with a limiting ohmic resistor
(R$_S$), probably associated with the conduction through a
filamentary YBCO which connects the interfacial zone to the bulk.
This equivalent circuit suggests the existence of 2 interfacial
regions, as depicted in Fig.~\ref{fig:circuito}.

Considering this possible equivalent circuit model suggested by the
$\gamma$
graphical representation of the I-V curves, the following equations can be derived: \\

\begin{equation}
\label{eq:P-F} i_{PF}(v_{PF}) = A v_{PF}~exp\left[ B \sqrt{v_{PF}},
\right]
\end{equation}

with

\begin{equation}
\label{eq:param} A = \tilde A_{PF}~\exp \left[ \frac{ -
{\overline\phi}_t}{T} \right] ~~;~~ B = \frac{q^{3/2}}{k_B T (\pi
\epsilon^{'}\epsilon_0 d)^{1/2}},
\end{equation}

\noindent where $i_{PF}$ and $v_{PF}$ are the current and the
voltage across the PF element, respectively. $\tilde A_{PF}$ is
associated with the geometric factor of the conducting path, the
electronic drift mobility ($\mu$) and the density of states in the
conduction band. $\overline\phi_t$ is the trap energy level (in K),
$q$ the electron's charge, $k_B$ the Boltzmann constant,
$\epsilon^{'}$ the real part of the relative dielectric constant of
the oxide, $\epsilon_0$ the permittivity of vacuum and $d$ the
distance associated with the voltage drop $v_{PF}$. Notice that $d$
is not necessarily equal to the distance between the voltage
contacts.

\noindent As $ v_{PF} = V - v_S = V_{Au} - v_S $, then:

\begin{equation}
\label{eq:I-V} I = {\rm{A}}({{\rm{V}}_{{\rm{Au}}}} -
I{R_S})exp\left[ {{\rm{B}}\sqrt {{{\rm{V}}_{{\rm{Au}}}} - I{R_S}} }
\right] + \frac{{{{\rm{V}}_{{\rm{Au}}}} - I{R_S}}}{{{R_P}}}
\end{equation}

\begin{figure}[!htb]
    \centering
    \includegraphics[scale=.6]{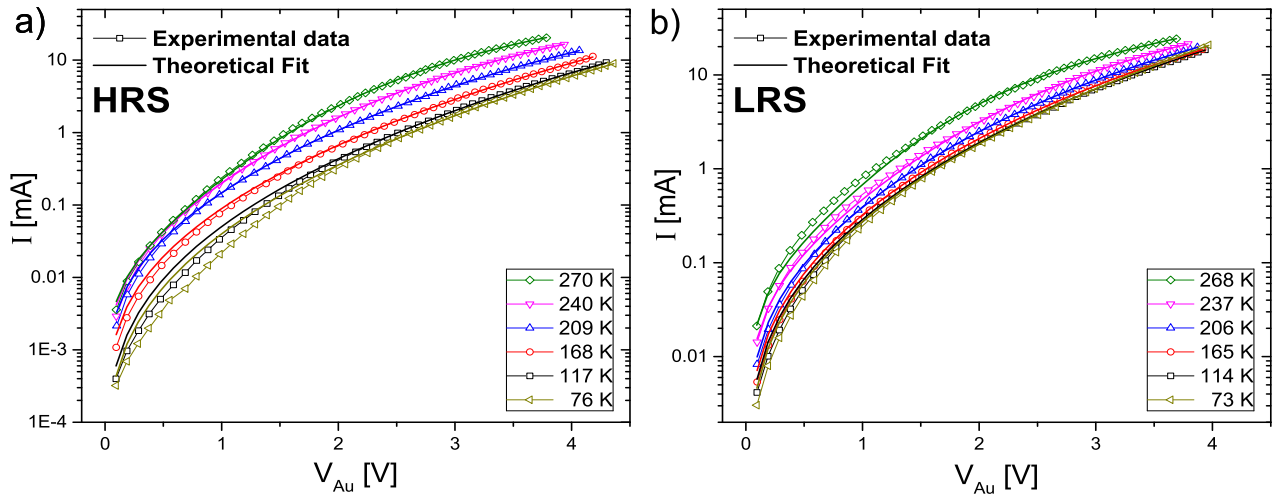}
    \caption{Experimental and theoretical (Eq.~\ref{eq:I-V}) I-V characteristics at
    different temperatures for the a) HRS and b) LRS.}
\label{fig:IVteoexpHRSyLRS}
\end{figure}

Eq.~\ref{eq:I-V} is an implicit equation that should be solved
numerically to fit the experimental data. As can be observed in
Fig.~\ref{fig:IVteoexpHRSyLRS} the experimental I-V characteristics
for the HRS and the LRS are very well reproduced by the theoretical
representation of the proposed circuit model (Eq.~\ref{eq:I-V}),
respectively (as well as the corresponding $\gamma$ curves shown in
Fig.~\ref{fig:gammaHRSyLRS}). A small deviation can be observed at
low currents and voltages, probably associated with the existence of
low thermoelectric voltages at the interfaces.

\begin{figure}[!htb]
    \centering
    \includegraphics[scale=.4]{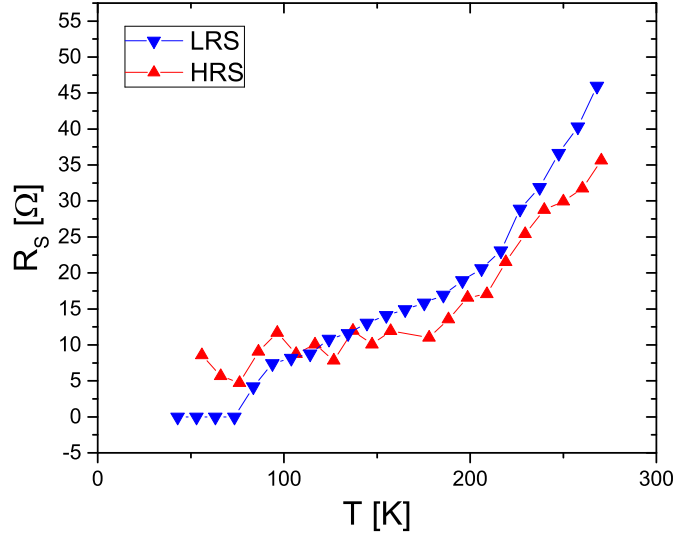}
    \caption{Series resistance R$_S$ as a function of temperature for the LRS and HRS.
    A drop to $\simeq$ 0 resistance can be observed in the LRS at a temperature close to
    the superconducting T$_c$ of the bulk YBCO, while this is not the case for the HRS.}
    \label{fig:paramRs}
\end{figure}

\begin{figure}[!htb]
    \centering
    \includegraphics[scale=.5]{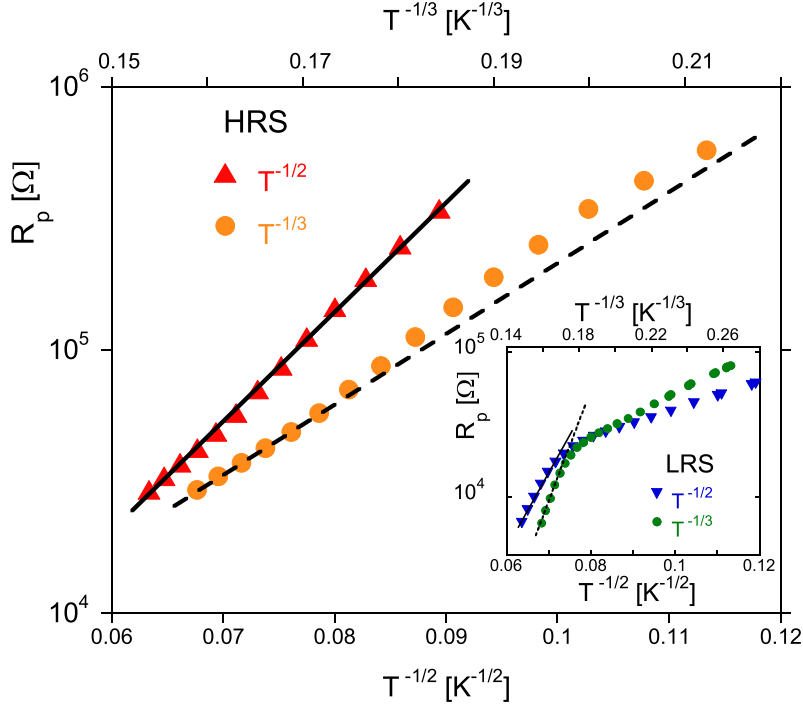}
    \caption{Parallel resistance R$_P$ as a function of T$^{-1/2}$ and T$^{-1/3}$ for the HRS.
        The inset shows the same plot for the LRS. Lines are guides to the eye.}
    \label{fig:paramRp}
\end{figure}

\begin{figure}[h]
    \centering
    \includegraphics[scale=.5]{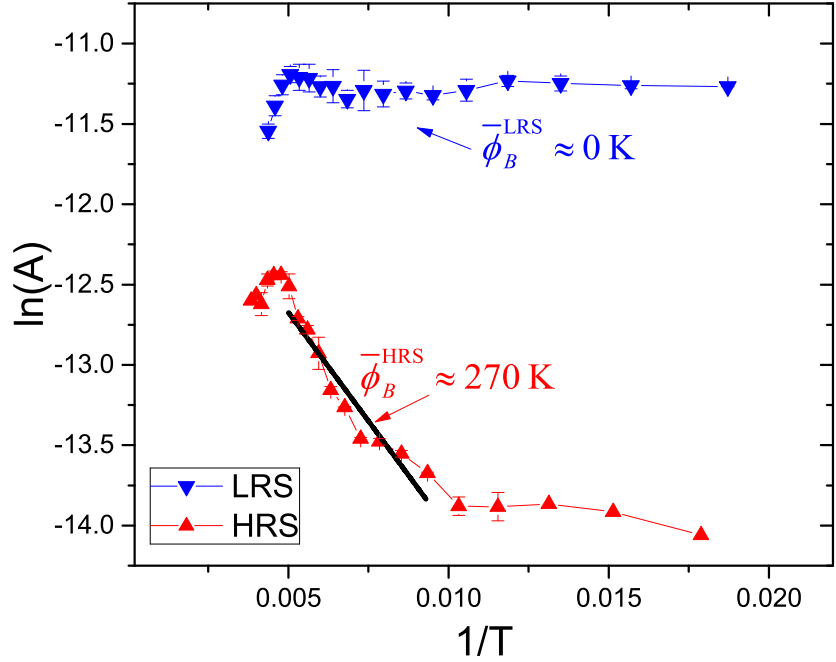}
    \caption{Semilogarithmic plot of the A parameter as a function of 1/T for the LRS and
    HRS. ${\overline\phi}_t$ is the mean energy of the carrier's
    traps, which is obtained from the linear dependence as stated
    from Eq.~\ref{eq:param} (see the line for the HRS data).}
    \label{fig:paramA}
\end{figure}

\begin{figure}[!htb]
    \centering
    \includegraphics[scale=.5]{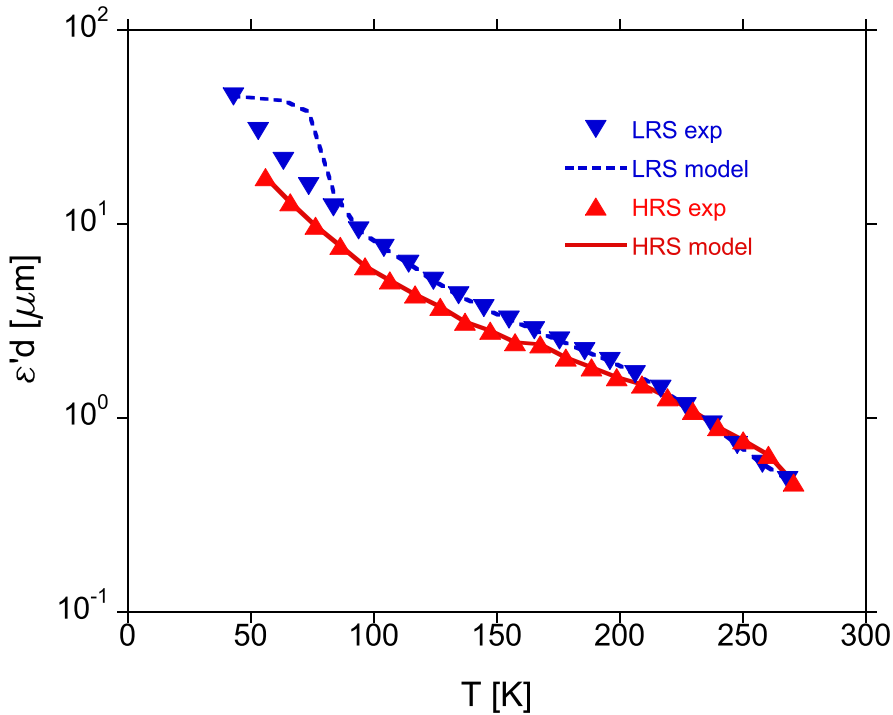}
    \caption{Temperature dependence of $\epsilon^{'}$d for the LRS and HRS, derived
    from the B fitting-parameter by using Eq.~\ref{eq:param}. Lines are fits obtained
    by considering the model described by Eq.~\ref{eq:epsilondeR}.}
\label{fig:paramEpsilon}
\end{figure}

In this way, the temperature dependence of the fitting parameters
$R_S$, $R_P$, $A$ and $B$ can be extracted, as shown in
Fig.~\ref{fig:paramRs}-\ref{fig:paramEpsilon}.

In the LRS, the series resistance R$_S$ mimics the bulk resistance
of the YBCO film, with even evidences of its superconducting
transition, as can be observed in Fig.~\ref{fig:paramRs}. As the
value of R$_S$ is at least one order of magnitude higher than
R$_{4W}$ ($\sim$ 0.1 $\Omega$), we can infer that this is a
consequence of a reduced geometric factor, as if the connection of
the interfacial zone to the bulk of the film is realized through
small conducting channels of near optimally doped YBCO through an
insulating matrix of oxygen depleted YBCO. In the HRS,  R$_S$ also
shows a metallic-like behavior, but no superconducting transition
can be detected, probably associated with the disappearance of the
percolation of the superconducting state through these still
metallic channels.

The existence of the mentioned insulating matrix is revealed by the
temperature dependence of R$_P$. Indeed, as can be observed in
Fig.~\ref{fig:paramRp}, the R$_P(T)$ data is better represented by
an Efros-Shklovskii variable range hopping (ES VRH) law (R$_P \sim
exp[(\frac{T_0}{T})^{1/2}]$) rather than by a 2D Mott localization
(R$_P \sim exp[(\frac{T_0}{T})^{1/3}]$), pointing to
YBa$_2$Cu$_3$O$_x$ with $x~\simeq$ 6 as the possible phase
associated with this particular electric
transport.~\cite{Matsushita87,Milliken00}

In fact, the non-linear conduction of YBa$_2$Cu$_3$O$_{6+\delta}$
with $\delta<$ 0.1 was already
reported~\cite{Matsushita87,Milliken00,Rey92} but its origin was not
addressed at that time. Here, in accordance to a previous study by
Schulman et al.~\cite{Schulman15}, we argue that the oxygen
vacancies no only induce disorder, which is at the basis of the
electronic localization, but also act as trapping centers,
determining an additional non-linear conduction channel based on the
PF effect.

By fitting the temperature dependence of the resistivity by the ES
VRH law, $T_0$ was obtained for the HRS and LRS ($\sim$ 9,000 K and
14,000 K, respectively). The localization
length~\cite{Efros84,Mott90} can then be determined as $\xi~=~2.8~
e^2/(\epsilon^{'} \epsilon_0~k_B T_0$). Assuming that
$\epsilon^{'}\sim 25$~\cite{Mannhart96} (see the discussion on
$\epsilon^{'}$ in the last section of this paper), we obtain that
$\xi^{HRS}~\simeq$~2.6 nm and $\xi^{LRS}~\simeq$~1.6 nm,
interestingly indicating that the insulating matrix in the LRS seems
to be more disordered than in the HRS. This indicates that when the
RS produces the LRS, not only a more conducting path is set but the
degree of disorder is increased, probably as a consequence of a noisy
distribution of oxygen vacancies.

This insulating phase is present for both the HRS and the LRS (see
the inset of Fig.~\ref{fig:paramRp}) but it can be observed that at
low temperatures there is a small increase of the conductance  (T
$\leq$ 180 K and T $\leq$ 100 K for the LRS and the HRS (not shown),
respectively, consistent with the presence of minor metallic zones.

The temperature dependence of the $A$ parameter, associated with the
PF conduction, is shown in Fig.~\ref{fig:paramA} for both HR and LR
states. Plotted as $Ln(A)$ vs T$^{-1}$, a linear dependence with a
negative slope can be observed for the HRS in a certain temperature
range. According to Eq.~\ref{eq:param} this slope represents the
mean energy of the PF traps (${\overline\phi}_t \simeq$ 270 K). This
value, which depends on the HRS attained, is in the order of
magnitude of the ones reported for ceramic YBCO/Au devices (700-1200
K).~\cite{Schulman15}

In the case of the LRS, the obtained value of ${\overline\phi}_t$ is
close to 0 K, although it should be non null as the non-linear
behavior is still observed for this state ($\gamma > 1$), indicating
the existence of shallow traps. Interestingly, in the HRS for T
$\lesssim$ 90 K, also ${\overline\phi}_t$ $\rightarrow$ 0 K, in
accordance to the appearance of the superconducting state in
optimally doped YBCO.

The extrapolation of the linear dependence for high temperatures,
which corresponds to $\tilde A_{PF}$, gives similar values for both
states, indicating that the main effect of the RS over the PF
conduction is related to the variation of the trap energy
${\overline\phi}_t $.

For T $\ge$ 220 K, both states show an anomalous behavior for the
$A$ parameter, which is reproduced both lowering and raising
temperature without hysteresis. This feature was observed in several
samples, sometimes making difficult for this temperature range to
find a good fit of the experimental IV curves, determining as a
consequence a noisy $A$ parameter. Interestingly, at 200-250 K,
partially oxygenated YBCO shows a phase transformation related to
the reordering of oxygen atoms in the Cu-O basal planes, which
softens its elastic properties.~\cite{Cannelli88,Cannelli92} As the
pulsing treatments that generate the RS modifies the number of
oxygen vacancies particularly in these planes, the appearance of
this transition would then be expected. Although this phase
transformation may generate misfitting domains due to non-elastic
accommodations, the fact that no hysteresis is observed in the $A$
parameter on cooling and heating indicates that the anomaly observed
may be related to a dynamical property which affects the carrier's
mobility. As this phase transformation was characterized as
relatively slow, evolving during several thermal cycles, our
transport measurements may result sensitive to this evolution in the
time scale of our experiments.

These results point out to a scenario of high inhomogeneity in terms
of oxygen concentration in the interfacial YBCO, with the
coexistence of insulating zones ($x~\simeq$ 6), dominated by a VRH
type transport and with the presence of PF traps, with a minority
metallic phase, which may even be superconducting ($x~\simeq$ 7).

In this context of metallic zones embedded in an insulating matrix,
the formation of thin (nanometric) capacitive layers in the
interfacial zone should be expected. This will lead to the
Maxwell-Wagner (MW) effect~\cite{VonHippel62}, which amplifies the
dielectric response of the device as well as its temperature
dependence, eventually giving rise to an extrinsic colossal
dielectric constant.~\cite{Lunkenheimer10} As this effect would not
be caused solely by the metallic contact between Au/YBCO but is
related to the inhomogeneities of the oxygen distribution in the
YBCO interfacial zone, it can be viewed as intrinsic to this complex
oxide, where the oxygen concentration can be varied spatially in an
inhomogeneous way, generating this particular mixture of zones with
different conductivities, acting as parallel-plate capacitors with
very small plate distances.

As previously shown~\cite{Lunkenheimer10}, this framework can be
described by a simple equivalent circuit consisting in two layers
connected in series: one for the interfacial zone ($i$) and the
other for the bulk ($B$). Both zones of the same area and with a
total width $d_T=d_i+d_B$. Each zone having a relative real part of
the dielectic constant $\epsilon_{i,B}$ and represented by a
conductance G$_{i,B}$ in parallel with a capacitance C$_{i,B}$. This
configuration leads to the following expressions for the frequency
($\omega$) dependence of the equivalent real dielectric constant
($\epsilon^{'}$), the relaxation time ($\tau$), and the capacitance
at high and low frequencies [$C(\infty)$ and $C(0)$, respectively]:

\begin{equation}
\label{eq:MW1} \epsilon^{'}(\omega) = \epsilon^{'}_{\infty} +
\frac{\epsilon^{'}_s-\epsilon^{'}_{\infty}}{1+(\omega~\tau)^2} ~~;~~
\tau = \frac{C_i + C_B}{G_i + G_B}
\end{equation}

\begin{equation}
\label{eq:MW2} C(\infty) = \frac{C_iC_B}{C_i+C_B}~~;~~C(0) =
\frac{C_iG_B^2+C_BG_i^2}{(G_i+G_B)^2}
\end{equation}

\noindent where $\epsilon_{s}$ and $\epsilon_{\infty}$ are the
static and the high frequency ($\omega$) limit of $\epsilon^{'}(w)$,
respectively.

It can be noted that if $G_B \gg G_i$ and $C_i \gg C_B$  (which is
the case if $i$ is an insulating dielectric layer) and the intrinsic
dielectric constant of each zone (i,B) varies slowly with
temperature, the following approximations can be done:

\begin{equation}
\label{eq:MW3} \tau(T)\simeq\frac{C_i}{G_B}\sim
G_B^{-1}(T)=R_B(T)~~;~~ C(\infty)\simeq C_B~~;~~C(0)\simeq C_i.
\end{equation}

\noindent If R$_B$(T) has a metallic-like dependence, then $\tau(T)$
decreases with decreasing T and concomitantly $\epsilon^{'}$,
measured at a fixed frequency, should increase (see
Eq.~\ref{eq:MW1}).

Fig.~\ref{fig:paramEpsilon} shows the temperature dependence of the
dielectric response for both resistance states, extracted from the
fitting parameter $B$, plotted as $\epsilon^{'}d$ (which is
essentially the temperature dependence of $\epsilon^{'}$ with the
assumption that $d$ is temperature independent). From these results,
it can be inferred that $\epsilon^{'}$ increases with decreasing
temperature. This behavior is contrary to the one reported for
oxygen poor YBCO ($x~\simeq$ 6)~\cite{Samara90,Rey92}, but fits very
well in the MW scenario that we have described. Notice that
$\epsilon^{'}~d$ is higher for the LRS than for the HRS. This result
is in full agreement with the expected from the MW effect insofar as
the interfacial zone is not short-circuited by the conductive phase.

This result is quite similar to the one reported for the
magnetic-field-dependent dielectric constant of
La$_{2/3}$Ca$_{1/3}$MnO$_3$~\cite{Rivas06}, where in the phase
separation scenario of the manganites, the magnetic field favors the
conversion of insulating regions into conducting ones, increasing
the mixture that potentiates the MW effect, until a maximum is
reached at the Curie temperature, when the metallic phase becomes
the main one.

In this context, the observed increase of $\epsilon^{'}$ with
decreasing temperature is associated with the temperature dependence
of the dielectric relaxation time [$\tau(T)$] for a fixed measuring
frequency ($\omega$), indicating that  $\epsilon^{'}$ is evolving
from $\epsilon^{'}_{\infty}$  to $\epsilon^{'}_s$ as stated in
Eq.~\ref{eq:MW1}. This means that at a certain temperature ($T_p$)
the experimental characteristic time equals the characteristic
relaxation time:

\begin{equation}
\label{eq:omega} \omega \simeq \frac{1}{C_i(T_P) R_B(T_p)},
\end{equation}
\noindent and by assuming that C$_i$ is slightly temperature
dependent in comparison with R$_B$, Eq.~\ref{eq:MW1} can be
rewritten as:
\begin{equation}
\label{eq:epsilondeR}
 \epsilon^{'}(T) \simeq \epsilon^{'}_{\infty} +
\frac{\epsilon^{'}_s-\epsilon^{'}_{\infty}}{1+[\frac{R_B(T)}{R_B(T_p)}]^2}.
\end{equation}

This two layer interface-equivalent circuit can be easily associated
with the one we propose to describe our IV characteristics: it would
only demand that the PF element acts as a resistor, which is the
case if a constant current is used for its measurements, and to
associate $R_B(T)$ with $R_S(T)$, that is easily fulfilled since the
bulk resistance of YBCO $R_b \ll R_S$. In this way, the temperature
dependence of $\epsilon^{'}d$ presented in
Fig.\ref{fig:paramEpsilon} can be fitted by considering
Eq.~\ref{eq:epsilondeR} with the $R_S(T)$ data presented in
Fig.~\ref{fig:paramRs} as $R_B(T)$ and by determining $\epsilon_sd$,
$\epsilon_{\infty}d$ and  $R_b(T_p)$ as fitting parameters.

An excellent fit is obtained (see the lines shown in
Fig.\ref{fig:paramEpsilon}). The model deviates from the
experimental parameter for the LRS when the superconducting
transition of $R^{LRS}_S(T)$ occurs, probably as a consequence of a
sudden increase in $C_i$, which contradicts the assumptions made of
its slight temperature dependence. The obtained parameters are
listed in Table 1.

\begin{table}[h]
    \caption{Parameters derived from fitting the temperature dependence of $\epsilon d$ by considering
    Eq.~\ref{eq:epsilondeR} and the obtained values of $R_b(T)$. }
    \vspace{0.4cm}
    \label{tab:ajustes}
    \begin{tabular}{||c||ccccccccc||}
        \hline  \hline
\textbf{State}      && \large{$\epsilon^{'}_{\infty}d$} && \large{$\epsilon^{'}_Sd$} && \large{$R_S(T_p)$}  && \large{$T_p$}  & \vspace{-2mm}\\
        && \textbf{($\mu$m)} && \textbf{($\mu$m)} && \textbf{($\Omega$)}  && \textbf{(K)}  & \\ \hline \hline
        \textbf{HRS} & &       0.1     & &    22.8      & &    4.7      & &  70  &        \\
        \textbf{LRS} & &       0.14       &&    47.3      &&     3.9       & &    60 &      \\ \hline \hline
    \end{tabular}
\end{table}

A crude estimation of $d$, and thus of $\epsilon^{'}_{\infty}$,
$\epsilon^{'}_S$ and $\epsilon^{'}_i$ can be made by considering the
approximate values of the equivalent capacitance at low and high
frequencies (see Eq.~\ref{eq:MW3}), by assuming that $d \sim d_i
\sim d_B \sim d_T/2$ and that $\epsilon^{'}_{B}$ can be associated
with the dielectric constant of YBCO, which ranges from 5 to 40,
depending on its
granularity.~\cite{Humlicek92,Mannhart96,Navacerrada12} In that case
it can be shown that:

\begin{equation}
\label{eq:d} d_T \simeq \frac{\epsilon^{'}_{\infty}
d}{\epsilon^{'}_{B}} ~~;~~\epsilon^{'}_{i} \simeq
\frac{\epsilon^{'}_{S} d}{d_T} ,
\end{equation}
\noindent which gives the possibility to estimate the mentioned
values for the HRS and the LRS, as summarized in Table 2.

\begin{table}[h]
    \caption{Estimation of $d_T$, $d$, $\epsilon^{'}_{\infty}$, $\epsilon^{'}_i$ and $\epsilon^{'}_S$
    derived from Eq.~\ref{eq:MW3} and ~\ref{eq:d} and the values reported in Table 1.}
    \label{tab:ajustes}
    \begin{tabular}{||c||ccccccccccc||}
        \hline  \hline
        \textbf{State}      && \large{$d_T$} && \large{$d$} && \large{$\epsilon^{'}_{\infty}$}  && \large{$\epsilon^{'}_{i}$}  && \large{$\epsilon^{'}_{S}$} & \vspace{-2mm}\\
        && \textbf{(nm)} && \textbf{(nm)} &&    && (x 10$^3$)  && (x 10$^3$)  & \\ \hline \hline
        \textbf{HRS} & &       2.5-20     & &    1.3-10      & &    10-80      &&  1-9  &&  2.3-18  &        \\
        \textbf{LRS} & &       3.5-28       &&   1.8-14      &&     10-80       &&  1.7-14  &&  3.4-27 &      \\ \hline \hline
    \end{tabular}
\end{table}

It can be observed that the width of the whole interface region
($d_T$) is of the order of 10 nm and increases when switching the
device from HRS to LRS. This value is in accordance to other
estimations in YBCO~\cite{Waskiewicz18} as well as in other RRAM
devices based on the oxygen diffusion mechanism.~\cite{Aoki14}

The large dielectric constant of the insulating layer
$\epsilon^{'}_{i}$ is in the range of the one reported for oxygen
depleted YBCO~\cite{Rey92}, although these published colossal values
can be affected by the extrinsic effect already mentioned and
associated with interfaces. In spite of the fact that we have shown
that the MW formalism gives an excellent description of the derived
dielectric behavior of the inhomogeneous interfacial region of YBCO,
the high values of $\epsilon^{'}_{i}$ can not be the result of the
formation of tiny parallel capacitors with an ordinary dielectric
constant in the 10-40 range, as this would require that their
distance between metallic plates to be $\leq 1 {\AA}$. In this way,
it should be envisioned an intrinsic dielectric constant of this
region much higher than the expected for regular dielectric oxides.

As a final remark, It should be noted that we have used a model that
propose an effective dielectric constant associated with the whole
interfacial zone, composed by two different material ($i, B$), which
reproduce the temperature dependence of the dielectric constant
extracted from the PF conduction. However, it is natural to
associate the interfacial region $i$ to the one where the PF
conduction is developed, so that the dielectric constant extracted
from the PF conduction (see Fig.~\ref{fig:paramEpsilon}) should be
more likely related to $\epsilon^{'}_i$ rather than to
$\epsilon^{'}$. The very good reproduction of its temperature
dependence by considering the two layer model leads us to think that
instead of having two well separated layers, these may be
distributed randomly in a set of equivalent multilayers, which
behave identically from the dielectric point of
view~\cite{VonHippel62} but are averaged to the MW derived values by
the electrical carriers in a relevant distance like the localization
length. Further studies are needed to determine the origin of the dielectric behavior of these interfaces.

\section{CONCLUSIONS}

The electrical transport properties of Au / YBCO films were studied
by measuring the temperature dependence of their I-V curves in the
HRS and the LRS. Its analysis reveals the existence of a complex
scenario, where an interfacial zone composed by two layers can be
distinguished from the bulk YBCO. In the first layer, the observed
electronic localization plus a non-linear behavior based on the PF
effect is consistent with the presence of YBCO6, while the extracted
temperature dependence of the dielectric response would also
require, in terms of the MW effect, the existence of a minority
metallic YBCO. This scenario, mainly generated by the
electric-field-assisted drift of oxygen produced to achieve the RS
of the device, depicts an inhomogeneous distribution of the oxygen
content in this interfacial zone, with a higher (lower) proportion
of YBCO7 in a matrix of YBCO6 in the LRS (HRS). In the second layer,
a series resistor associated with a percolating metallic-like
conductive YBCO phase ensures a conducting path through small
channels with the bulk YBCO zone. The dielectric constant extracted
from the PF conduction follows a Debye relaxation law, consistent
with a MW scenario, but requiring a much higher intrinsic dielectric
constant for the insulating layer, outside the usual value for
oxides.

\section{ACKNOWLEDGEMENTS}
We would like to acknowledge financial support by CONICET Grant PIP
11220150100653CO, PICT 2017-0984 and UBACyT 20020170100284BA
(2018-2020). Jenny and Antti Wihuri Foundation is also acknowledged
for financial support. We are indebted to A. Schulman, M. Boudard
and C. Jim\'enez for fruitful discussions in an early stage of this
research. We thank D. Gim\'enez, E. P\'erez Wodtke and D.
Rodr\'{\i}guez Melgarejo for their technical assistance.


\end{document}